# Stabilizing van der Waals NbOI$_2$ by SiO$_2$ encapsulation for Photonic Applications

Gia Quyet Ngo,[1,2*] Fatemeh Abtahi,[1] Jakub Regner,[3] Hossein Esfandiar,[2] Peter Munzert,[2] Jan Plutnar,[3] Zdeněk Sofer,[3] Falk Eilenberger,[1,2,4] and Sebastian W. Schmitt[1,2]

[1]*Institute of Applied Physics, Abbe Center of Photonics, Friedrich Schiller University Jena, Albert-Einstein-Str. 15, 07745 Jena, Germany*

[2]*Fraunhofer-Institute for Applied Optics and Precision Engineering IOF, Albert-Einstein-Str. 7, 07745 Jena, Germany*

[3]*Department of Inorganic Chemistry, University of Chemistry and Technology Prague, Technicka 5, 16628 Prague 6, Czech Republic*

[4]*Max Planck School of Photonics, Germany*

*quyet.ngo@uni-jena.de

**Niobium oxide diiodide (NbOI$_2$) is an emerging material for photonics and electronics, distinguished by its exceptional second-order nonlinearity and pronounced in-plane ferroelectricity, both originating from its highly anisotropic ABC-stacked crystal structure. Its broken inversion symmetry enables its optical nonlinear efficiency to scale with thickness, making multilayer NbOI$_2$ highly promising for nonlinear frequency conversion like second harmonic generation or and spontaneous parametric down-conversion in bulk or waveguides. However, under ambient conditions NbOI$_2$ degrades into an amorphous oxide within weeks, severely diminishing its nonlinear response. To overcome this, we investigate SiO$_2$ encapsulation via physical vapor deposition to protect NbOI$_2$ multilayers from environmental degradation. Our systematic study reveals that encapsulation preserves structural integrity and nonlinear optical performance, establishing NbOI$_2$ as a stable candidate for heterogeneous integration in foundry-compatible photonic platforms and quantum technologies.**

Introduction

Two-dimensional (2D) materials have been widely explored for their strong linear and nonlinear interaction with light[1–8], high sensitivity to the surrounding environment[9–14], and excellent electrical properties[15–18]. Moreover, integrating 2D materials with optical waveguides has been shown to significantly enhance their optical response[19–25]. However, the challenging optical coupling of these semiconducting materials to foundry-processed photonic platforms[26] and the lack of foundry-process-compatible protection against the surrounding environment still hinders their applicability in back-end optics and photonics devices[27,28]. Recently, niobium oxide halides, such as $NbOI_2$, have garnered significant interest because of their optical and electronic properties, including broken inversion symmetry, large nonlinearity, and piezo- and ferroelectricity[29–37] that can be further enhanced by an external electric field[30]. This material group also shows potential in quantum technology, as demonstrated by the realization of spontaneous parametric down-conversion in films of only 50 nm thickness[31]. Yet, $NbOI_2$ is sensitive to ambient conditions, with the oxidation process turning the crystals from a crystalline to an amorphous structure[31,35,36]. As a result, the second-harmonic generation (SHG) drops dramatically and limits their applicability in long-term use. An attempt to protect $NbOI_2$ multilayers by a manual transfer of hexagonal boron nitride has been demonstrated[30,35]. However, this approach lacks precise thickness control, and it is not a simple and scalable integration using established coating techniques. Hence, reliable protection of $NbOI_2$ with a PIC-compatible deposition technique is highly desired for future heterogeneous photonic systems.

In this work, we demonstrate a scalable approach to encapsulate $NbOI_2$. Multilayers were exfoliated and mechanically transferred to $SiO_2$ substrates for encapsulation and guided-wave studies. We used a physical vapor deposition (PVD) process to grow a 75 nm thin film of $SiO_2$ to encapsulate $NbOI_2$. Although oxide coatings have been used to preserve some 2D materials[38–42] previous works simply focused on the electrical properties. Here, we studied the structure and nonlinear optical properties of $NbOI_2$ after the encapsulation process with $SiO_2$. Our goal was to keep the oxide films as thin as possible and fully cover $NbOI_2$ while minimizing any impact on their inherent physical properties. Optical microscopy, Raman spectroscopy and SHG have been used to characterize the encapsulated and unencapsulated reference $NbOI_2$. We observe that the crystal structure and the geometry of the $NbOI_2$ multilayers were well protected with 75 nm $SiO_2$ coating, while the reference sample was significantly deteriorated after 31 days in ambient conditions. The encapsulation study enables the exploration of engineered heterostructures of $NbOI_2$ and thin oxide films, toward potential applications in photonics, including bulk or hybrid waveguide systems, micro-resonators, and distributed Bragg reflectors. Our findings apply to nonlinear optics based on 2D materials and contribute to the evolution of future heterogeneous integrated photonic circuits.

**Results and Discussion**

**Crystal structure**

Millimetre-sized $NbOI_2$ single crystals were synthesized by chemical vapor transport (CVT) using niobium, niobium pentoxide, and iodine in a quartz glass ampoule (Details in the Methods section). After synthesis, the $NbOI_2$ crystal, which belongs to the chiral and polar space group $C_2$[29,30], shows atomic layers stacking along the a-axis by the weak van der Waals force, as shown in Figure 1a. The one-dimensional Peierls distortion between Nb atoms occurs along the c-axis, leading to non-zero polarization along the crystallographic b-axis[43], as indicated in Fig. 1a. Each layer consists of $NbO_2I_4$ octahedra and has a thickness of 0.73 nm[30]. This ABC stacking structure induces a broken inversion symmetry for each layer and supports substantial optical nonlinearity. To characterize the as-grown $NbOI_2$ crystal, Fig. 1b shows a scanning electron microscopy (SEM) image revealing multiple layers.

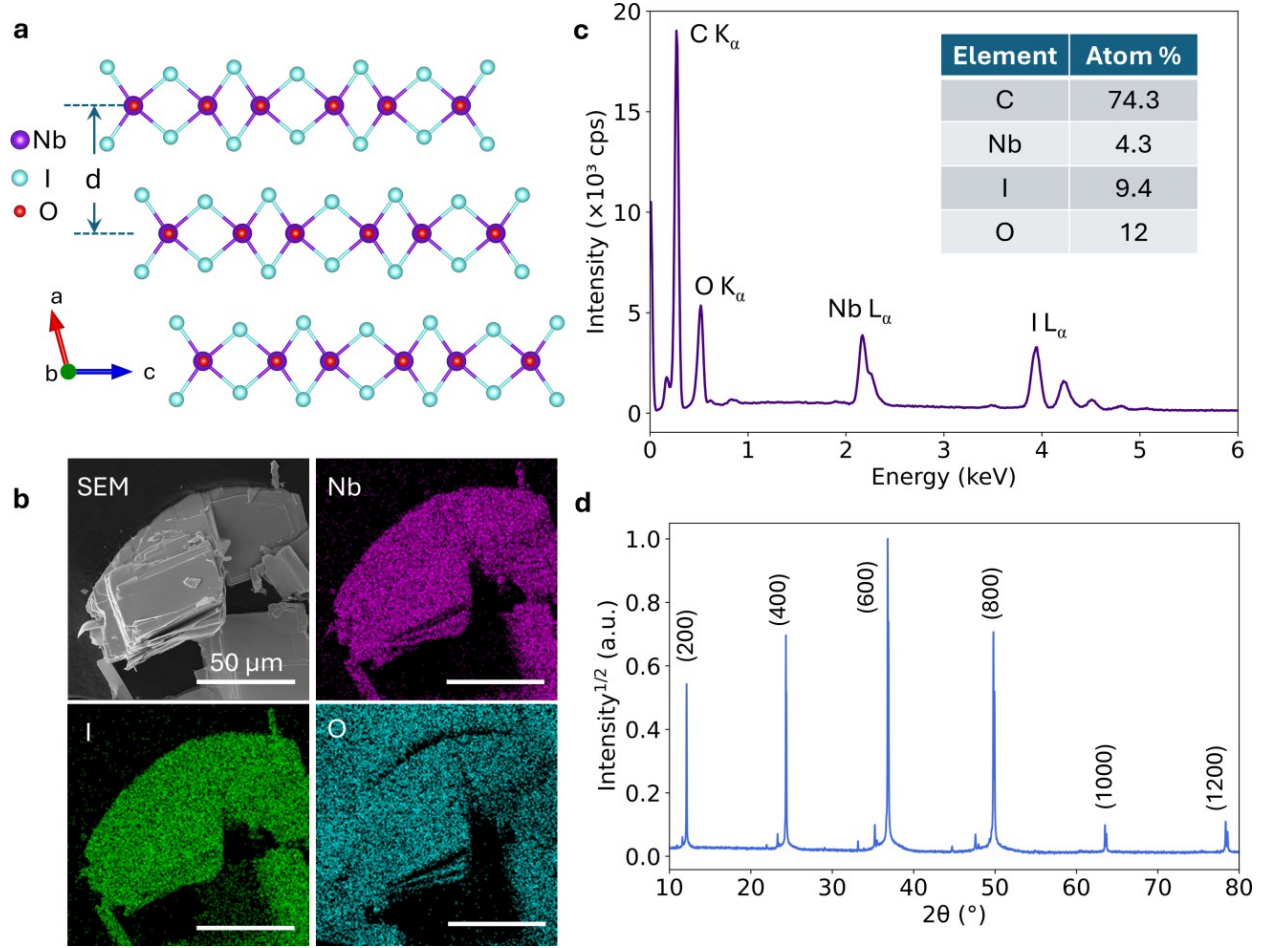

**Figure 1.** Crystal structure and material characterization. **a**, Side view of trilayer $NbOI_2$ viewed along the non-polar c-axis and stacked along the a-axis. The thickness d of a monolayer is 0.73 nm. **b**, SEM image and corresponding EDS mapping of van der Waals $NbOI_2$. The scale bar is identical for all subfigures. **c**, EDS spectrum of the $NbOI_2$ crystal and the calculated atomic percentage. **d**, XRD pattern of the $NbOI_2$ crystal in **b** with six difraction peaks. The square root of intensity is used to better observe the weak peaks.

Next, quantitative energy-dispersive X-ray spectroscopy (EDS) was employed to map the distribution of Nb, I, and O atoms. Further analysis of the peak areas using the EDS spectrum in Fig. 1c indicated the atomic percentages of the main components. The atomic ratio between Nb and I is nearly 1:2, matching the stoichiometric ratio of $NbOI_2$. The excess presence of O atoms and the high concentration of C atoms originate from the carbon tape that is used to mount the specimen for analysis, thus preventing us from obtaining the correct stoichiometric ratio for O. The X-ray diffraction pattern (XRD) in Fig. 1d shows six peaks for $NbOI_2$ at 12.10° (200), 24.32°

(400), 36.82° (600), 49.82° (800), 63.54° (1000), and 78.34° (1200). The XRD pattern of the synthesized $NbOI_2$ was then compared with literature[32,35,44], confirming the high crystallinity and correct ferroelectric $C_2$ phase. The diffraction pattern exhibits a series of (h00) reflections, namely (200), (400), (600), (800), (1000), and (1200). The presence of solely (h00) peaks demonstrates that (100) planes are parallel to the substrate, with the a-axis oriented out-of-plane while the b- and c-axes are in-plane. This confirms that the crystal is stacked along the a-axis.

**Optical microscopy**

Figure 2 shows optical microscopy (OM) images of mechanically exfoliated $NbOI_2$ flakes on fused silica wafers with a size ranging from 100 to 200 μm. The unencapsulated reference flakes are presented in Figs. 2a-d with the thickness determined to be in the range of 30-240 nm, while the encapsulated flake is shown in Figs. 2e-h, with the thickness determined to be in the range of 40-360 nm (Details in Supplement S1-S2).

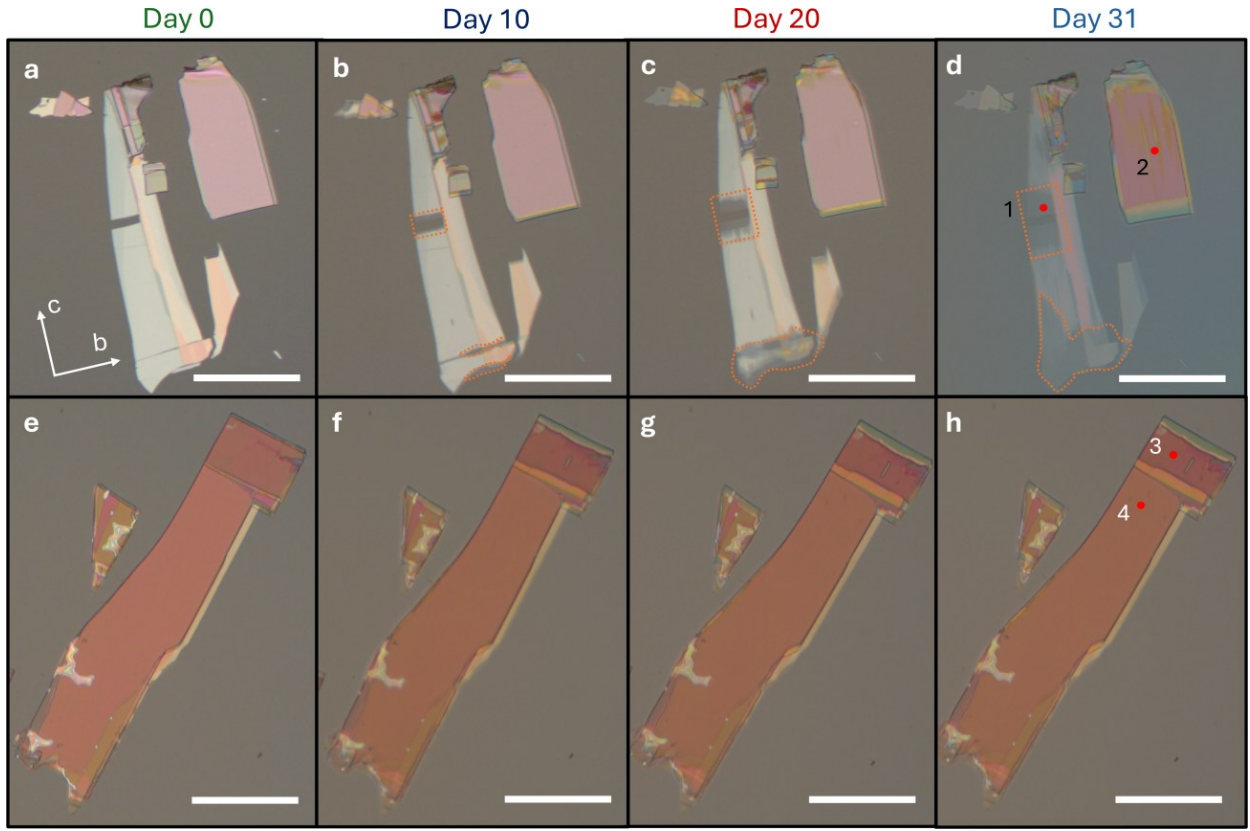

**Figure 2.** Optical microscopy images of $NbOI_2$ flakes, stored in air over time. **a-d**, Unencapsulated $NbOI_2$ measured at day 0, day 10, day 20, and day 31, respectively. Points 1 and 2 indicated the positions to measure Raman spectra. Selected degraded areas are marked with dash-dotted lines to guide the eye. b and c are crystallographic axes of the sample. **e-h**, $NbOI_2$ was encapsulated with $SiO_2$ thin film and measured at day 0, day 10, day 20, and day 31 after the encapsulation, respectively. Points 3 and 4 indicated the positions to measure Raman spectra. Scale bars are 50 μm.

We monitor the shape of both flakes within 31 days to observe the impact of ambient conditions on the crystal morphology. Both flakes were stored in an identical chamber at air conditions with 60% relative humidity. The unencapsulated reference samples showed degradation after 10 days. Shrinkage of the flakes occurs along the crystallographic c-axis towards the centre, consistent with previous works[31,35,36]. Degraded regions are delineated by dash-dotted lines for clarity, revealing a pronounced progression of degradation between day 10 and day 31. Due to the oxidation process, $NbOI_2$ interacts with $H_2O$ and $O_2$ in the environment, forming transparent amorphous $Nb_2O_5$, which has virtually zero SHG. We confirm the formation of amorphous material in the sections below using Raman spectroscopy and second-harmonic generation (SHG), both of which are highly sensitive to the material's structural integrity.

To decelerate the degradation of $NbOI_2$ from the environment, we used pure $SiO_2$ (99,99%) evaporated in a high vacuum condition with a constant deposition rate of 0.4 nm/s by a PVD machine to deposit a 75 nm $SiO_2$ thin film on top of $NbOI_2$ (Details in Methods section). The morphology of the encapsulated flake was unchanged after the deposition process, as displayed in Figs. 2e-h. After 31 days in ambient conditions, the encapsulated sample retains its shape. Using the same PVD technique, the $Al_2O_3$ encapsulation of exfoliated $NbOI_2$ also provides efficient protection from the environment. The OM images of the $Al_2O_3$-encapsulated sample are presented in Supplement S3.

**X-ray photoelectron spectroscopy**

The surface composition of an unencapsulated sample after a few days was studied using X-ray photoelectron spectroscopy (XPS) with a SPECS spectrometer equipped with a monochromatic Al Kα X-ray source (1486.7 eV) and a hemispherical electron analyzer Phoibos 150. The survey spectrum was recorded with $E_p$ set to 100 eV, and the high-resolution spectra of the core lines with $E_p$ set to 40 eV. The base chamber pressure during the acquisitions was at $10^{-9}$ mbar or lower.

The survey spectrum of the $NbOI_2$ crystal in Fig. 3a reveals the presence of the expected elements in the expected 1:1:2 ratio by integrating the area under an element-specific peak and applying a correction to the intensity by relative sensitivity factors. Due to the manipulation of the sample in the ambient atmosphere, a small carbon signal corresponding to the adventitious carbon contamination is visible at approximately 284 eV. High resolution spectra of the O 1s, I 3d and Nb 3d core lines were recorded and presented in Figs. 3b-d, respectively. There is one pair of peaks at 620.3 and 631.7 eV, corresponding to the 5/2 and 3/2 fractions of the I 3d electrons, visible in the spectrum shown in Fig. 3c. Deconvolution of the high-resolution spectrum of the O 1s electrons suggests the presence of electrons with two different energies at 531.0 and 532.4 eV assigned to originate in $NbOI_2$ and $Nb_2O_5$, respectively. The presence of $Nb_2O_5$ is also visible in the high-resolution spectrum of the Nb 3d electrons in Fig. 3d; deconvolution of the data suggests a pair of signals originating in $NbOI_2$ (205.8 and 208.6 eV for the 5/2 and 3/2 fraction, respectively) and $Nb_2O_5$ (208.0 and 211.0 eV). We presume that $Nb_2O_5$ is formed on the surface of the crystal by the action of atmospheric oxygen during the sample's exposure to the atmospheric conditions, which is consistent with our observation in the optical microscopy section.

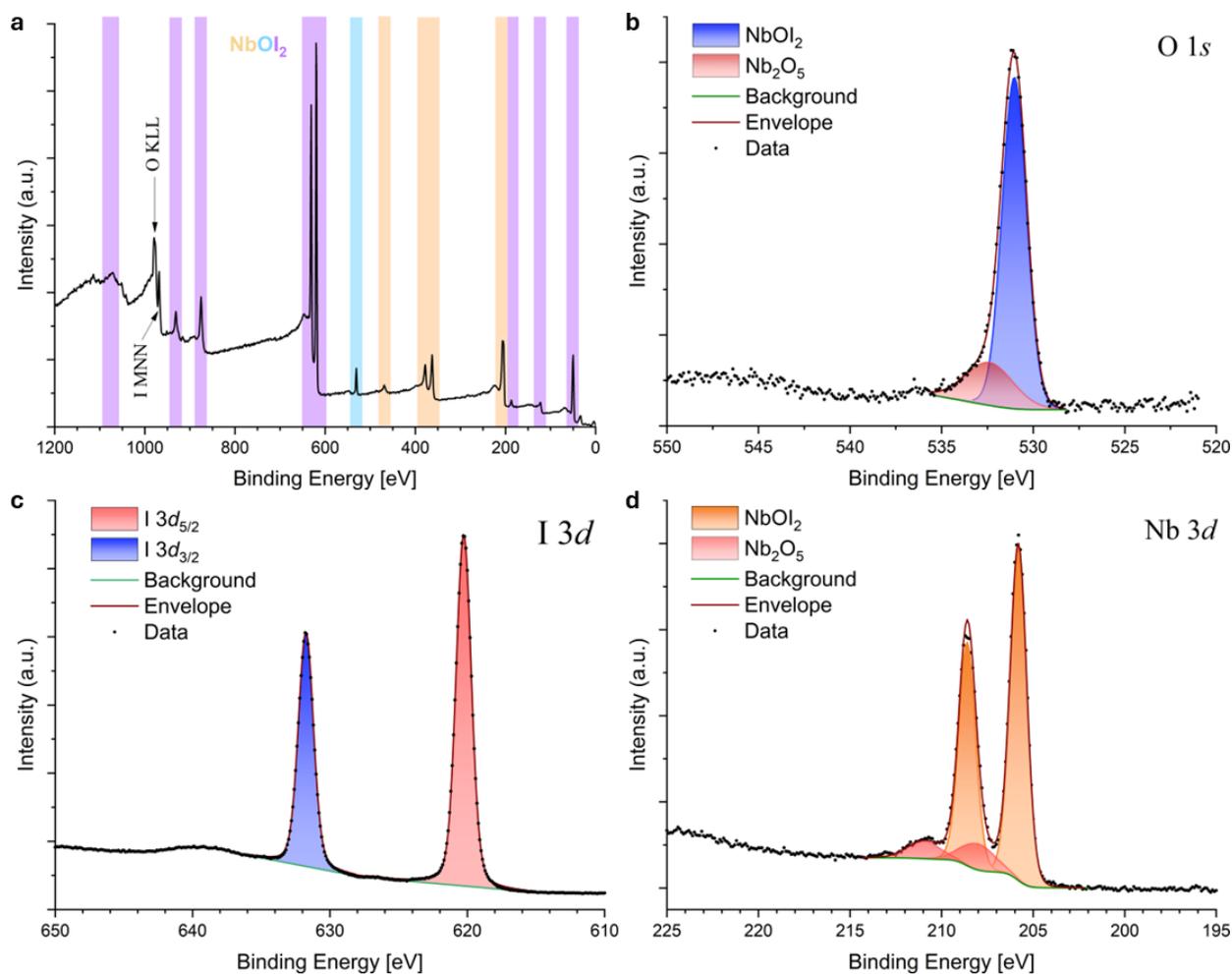

**Figure 3**. X-ray photoelectron spectroscopy of as-grown $NbOI_2$. **a**, A survey spectrum of the $NbOI_2$ crystal to reveal the stoicheometric ratio. **b-d**, High resolution spectra of O 1s, I 3d and Nb 3d.

**Raman spectroscopy**

Figure 2d and Fig. 2h show the positions where we monitored Raman spectroscopy over time. The measured points are labelled and marked for ease of presentation as red dots. Raman spectroscopy was examined at the identical position from day 0 to day 31. Fig. 4a shows the normalized Raman spectra of position 1 on the unencapsulated reference $NbOI_2$ sample marked in Fig. 2d. From Fig. 2d, one can see that position 1 is in the degraded area. At day 0, Raman spectra exhibit four distinct modes. Peaks $P_1$ at 104 cm$^{-1}$, $P_2$ at 207 cm$^{-1}$, $P_3$ at 271 cm$^{-1}$, and $P_5$ at 612 cm$^{-1}$ match well with reported values from literature for $NbOI_2$[29,35,36]. Raman spectra recorded from day 10 to day 31 show no Raman peaks at that position, indicating the oxidation of the $NbOI_2$. Fig. 4b shows the normalized Raman spectra of position 2, located on another unencapsulated flake with a higher thickness (see Supplement S1). Raman spectra show four distinct modes $P_1$, $P_2$, $P_3$ and $P_5$ from day 0 to day 31, like Fig. 4a.

Figure 4c shows the normalized Raman spectra of position 3 on the encapsulated sample. The Raman spectra exhibit five distinct modes. Peaks $P_1$ at 104 cm$^{-1}$, $P_2$ at 207 cm$^{-1}$, $P_3$ at 271 cm$^{-1}$, $P_4$ at 510 cm$^{-1}$, and $P_5$ at 612 cm$^{-1}$ similar to the unencapsulated sample. The peak $P_4$ is weaker and broader than the other peaks. Raman peaks $P_1$, $P_2$, $P_3$, and $P_5$ represent $A_g$ vibration, whereas $P_4$ links to $B_g$ vibration[29,36]. The position and intensity of the Raman peaks show almost no change over time. This indicates that the PVD coating does not alter the crystal structure and can protect them from the environment. Fig. 4d shows normalized Raman spectra of position 4 marked in Fig. 2h from day 0 to day 31. The data from day 0 is taken from position 3 for ease of presentation because position 4 was not measured from the beginning. Nonetheless, we observed similar characteristics as obtained at position 3. We also measured the Raman spectrum of $Al_2O_3$-encapsulated $NbOI_2$ as presented in Supplement S3. The intrinsic Raman peaks of $NbOI_2$ are almost unchanged between the two encapsulation oxides. However, the intensity of Raman peaks from the $Al_2O_3$-encapsulated sample is more homogeneous than in the case of $SiO_2$ encapsulation, which can be attributed to substrate-induced optical interference[45].

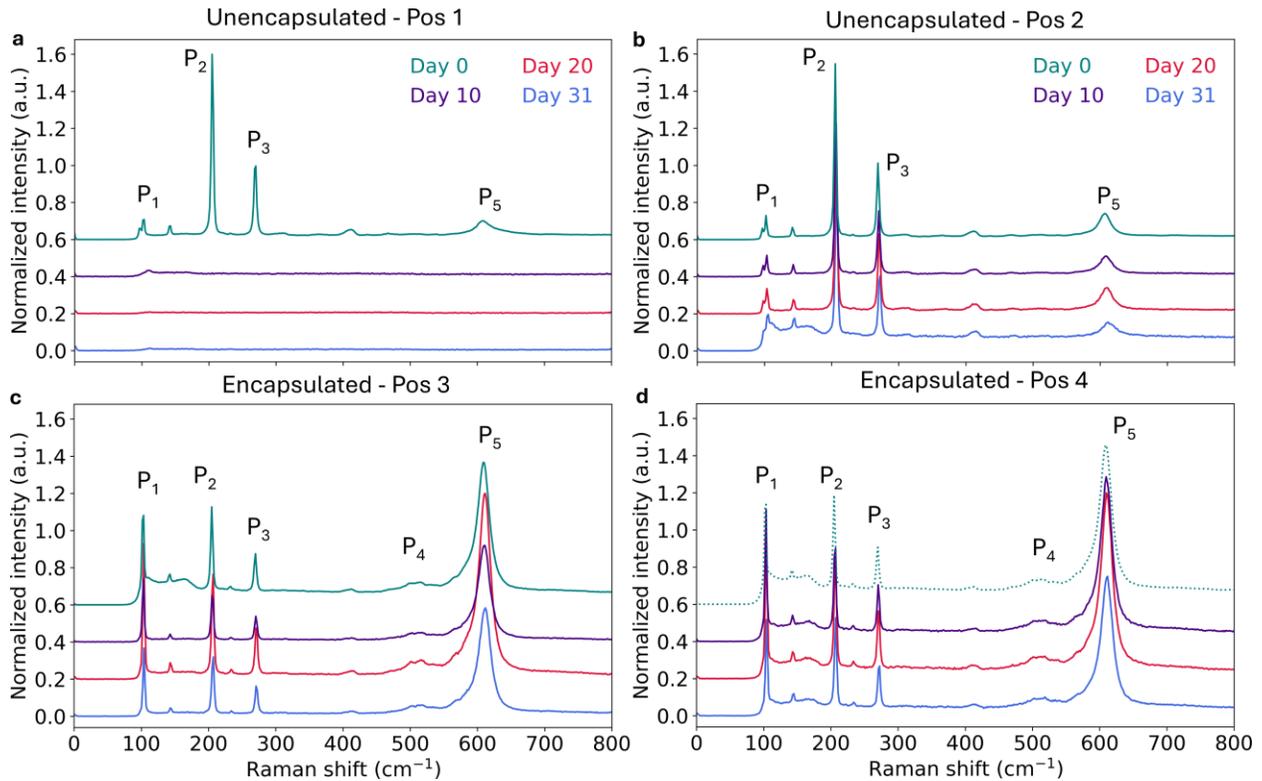

**Figure 4.** Raman spectroscopy of unencapsulated $NbOI_2$ and encapsulated samples, measured at day 0 to day 31. **a**, Normalized Raman spectra of position 1 on the unencapsulated sample indicated in Fig. 2d. **b**, Normalized Raman spectra of position 2 on the unencapsulated sample indicated in Fig. 2d. **c**, Normalized Raman spectra of position 3 on the encapsulated sample indicated in Fig. 2h. **d**, Normalized Raman spectra of position 4 on the encapsulated sample indicated in Fig. 2h. The dotted line of Day 0 is taken from position 3 for ease of presentation. Legends in **c** and **d** are identical to **a** and **b**.

The position of Raman peaks measured on the unencapsulated sample is similar to ones measured on the encapsulated sample. With the unencapsulated sample, peak $P_4$ is hardly seen, and the relative intensity ratios of $P_1/P_2$ and $P_5/P_2$ are much lower than in the case of encapsulated samples. We attribute this to the influence of the dielectric environment, the effect of degradation and the lower thickness of the unencapsulated sample. Overall, Raman spectroscopy has demonstrated the drastic change in the crystal structure of degraded $NbOI_2$.

**Second-harmonic generation**

Figure 5 shows SHG maps of reference $NbOI_2$ samples and samples encapsulated by $SiO_2$. A fs pulsed laser with vertical input polarization was used for both samples and kept unchanged for all experiments. The excitation laser at 1550 nm was focused on the samples using an NIR 50x objective. The signal SHG was collected by another NIR 50x objective in transmission configuration and imaged on a sCMOS camera after passing through a set of filters (see Methods for details). No output analyzer was used before the camera in this investigation. A scan range of 90 × 90 $\mu m^2$ was used to record the transmitted second harmonic emission from $NbOI_2$. Figs. 5a-d show the SHG maps of the unencapsulated reference $NbOI_2$ at day 0, day 10, day 20, and day 31, respectively. The sample exhibits degradation over time from day 0 to day 31, as indicated by white boxes for a thin flake (thickness of 30-60 nm in Supplement S1). Some parts of the thin flake started to shrink along the c-axis with almost zero SHG. As discussed in the previous parts, the oxidation process of $NbOI_2$ turns the degraded area into amorphous $Nb_2O_5$, which does not support SHG. The degraded area expanded over time and is noticeable from day 10 based on the shape and SHG intensity of the sample. For the flake with a thickness of around 220 nm (see Supplement S1 for topography), the change of total area is also noticeable, indicated by the green box. The length of the flake decreased from 79.2 $\mu m$ at day 0 to 68.2 $\mu m$ at day 31, along the c-axis. It demonstrates the sensitivity of SHG in probing the degradation of $NbOI_2$ semiconductors. In comparison to OM, in which, e.g., the thicker unencapsulated flake in Figs. 2a-d shows hardly any change in geometrical shape. SHG on this flake also indicates that flakes not only degraded along the c-axis, but also from the top surfaces. This is visible by the lines with dimmed SHG in the center of the thicker flake in Fig. 5d, which in turn extend along the in-plane c-axis.

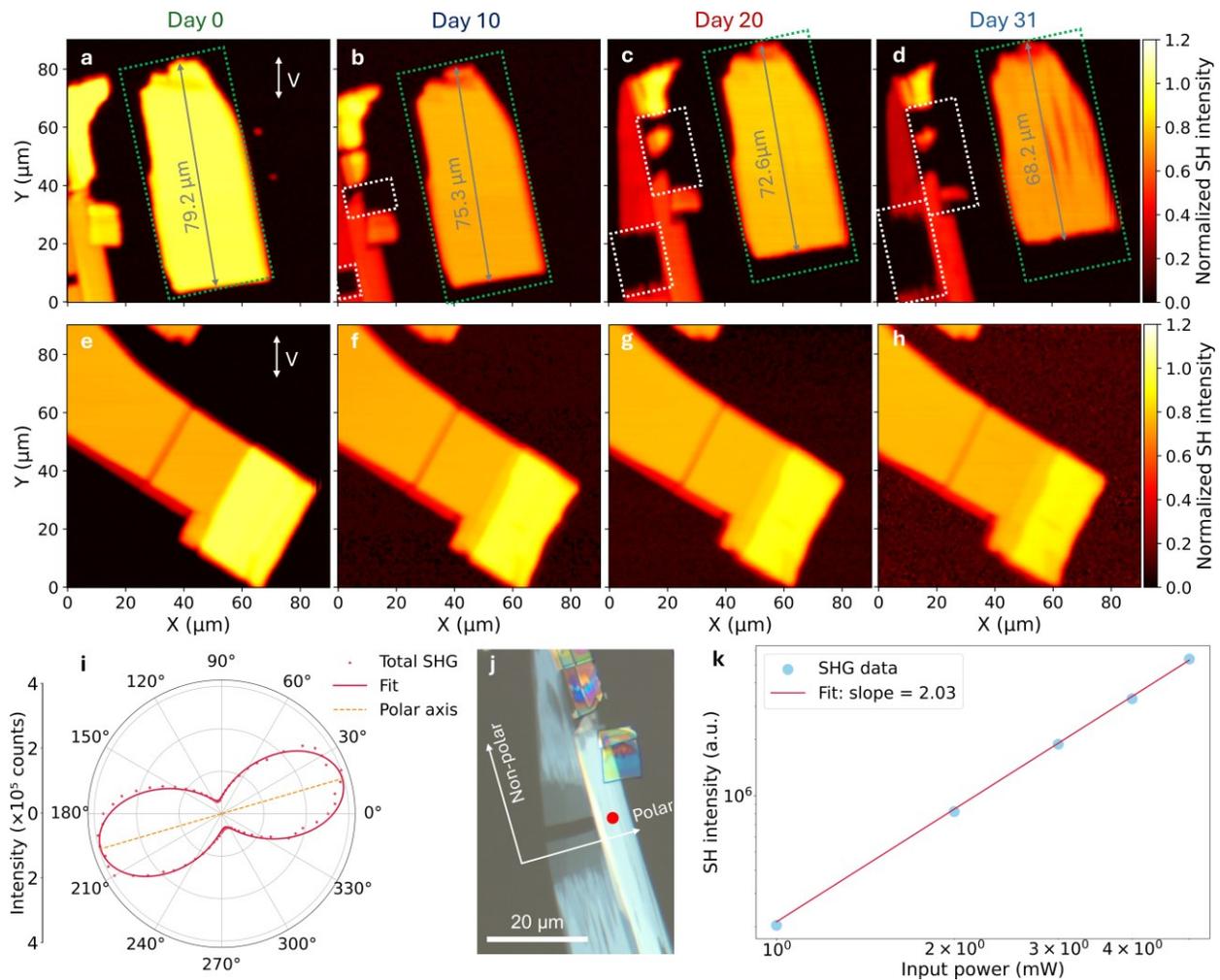

**Figure 5.** SHG microscopy of NbOI$_2$. **a-d**, SHG mapping of the unencapsulated reference NbOI$_2$ samples, measured at day 0, day 10, day 20, and day 31, respectively. **e-h**, SHG maps of the NbOI$_2$ encapsulated with SiO$_2$ thin film, measured at day 0, day 10, day 20, and day 31, respectively. The excitation wavelength is 1550 nm. V is vertical input polarization. White boxes indicate the degradation area. The green box indicates the original boundary of an unencapsulated flake with a thickness of 200 nm. **i**, Polarization-dependent SHG of the unencapsulated reference NbOI$_2$ samples with total SHG intensity. The red dots represent the measured data, while the solid lobes represent the theoretical predictions of the SHG intensity as a function of the input polarization. **j**, Optical microscopy image of the measured unencapsulated reference NbOI$_2$ samples in **i** with indicated polar and non-polar axes. Red dot indicates the measured position in **i**. **k**, Power dependence of the unencapsulated reference NbOI$_2$ sample measured with the excitation pump of 1550 nm.

Figs. 5e-h show the SHG maps of the encapsulated NbOI$_2$ at day 0, day 10, day 20, and day 31, respectively. The encapsulated flake has a thickness ranging from 40 to 360 nm. With the protection of a 75 nm SiO$_2$ thin film, the average SHG yield of the encapsulated sample is almost unchanged from day 0 to day 31. The sample's overall shape is retained.

Fig. 5i presents the polarization-dependent SHG intensity emitted from the unencapsulated reference $NbOI_2$ samples in the transmission configuration using the identical experimental setup. The measured position is marked as the red dot in Fig. 5j. The excitation wavelength was fixed at 1550 nm, while a motorized half-wave plate placed after a linear polarizer was used to control the input polarization, allowing full 360° rotation around the sample with a step of 5°. The starting polarization angle corresponds to the horizontal axis of the lab system. To collect the total SHG intensity, there was no analyzer at the output. The recorded total SHG intensity shows anisotropic characteristics with the maximum SHG along the polar axis of the crystal, whereas the non-polar axis exhibits the minimum SHG. This in-plane anisotropic response is a strong indicator to reveal the crystal axes of the material, which can be predicted from the OM images of degraded samples, and matches well with other works[30,31]. A typical power-dependent measurement was performed for the unencapsulated reference $NbOI_2$ sample, as shown in Fig. 5k. The power dependence of the recorded emission has a slope of 2.03, confirming its SHG nature.

**Conclusions**

In summary, the thin encapsulation of synthesized van der Waals $NbOI_2$ with $SiO_2$, studied through OM, Raman, and SHG measurements, demonstrates the protection of this material from the surrounding environment, while retaining its intrinsic optical properties. OM images reveal the degradation of $NbOI_2$ at ambient conditions, developing along the c-axis, and the degraded area becomes transparent. Raman spectroscopy results confirm that the degraded area does not show distinct Raman peaks as observed in pristine $NbOI_2$ flakes that are considered amorphous. The Raman peak positions are independent of the thickness; however, the intensity of individual peaks varies with different thicknesses. Besides Raman spectroscopy, SHG spectroscopy is also sensitive to slightly oxidized samples due to the strong dependence of SHG intensity on thickness and crystal structure. The oxidized area shows virtually zero SHG, which is expected by the nature of the symmetric amorphous material. Also, SHG maps can reveal the gradual oxidation happening with $NbOI_2$. The substantial nonlinearity of $NbOI_2$ may find interest in ultrafast optical modulators, nonlinear photonics, wave-guided systems, and quantum technologies. Our scalable encapsulation approach has demonstrated efficiency in maintaining the chemical and optical properties of the $NbOI_2$ and can pave the way towards integrating $NbOI_2$ into next-generation photonics circuits. These findings can apply to a range of oxides, other 2D materials, and various photonic device fabrication techniques.

**Methods**

**Material growth.** $NbOI_2$ bulk was made by chemical vapor transport (CVT) using niobium, niobium pentoxide, and iodine in a quartz glass ampoule. For the synthesis, we used Nb (-100 mesh, 99.9%, Strem, USA), $Nb_2O_5$ (-100 mesh, 99.9%, Strem, USA) and iodine (granules, 99.9%, Fisher Scientific, Germany) in a stoichiometric ratio corresponding to 15g of $NbOI_2$ with 0.3 g excess of iodine. The quartz ampoule (30 x 250 mm) with precursors was melted and sealed under high vacuum (<1x10$^{-3}$ Pa using an oil diffusion pump and liquid nitrogen cold trap) with an oxygen-hydrogen welding torch. The ampoule was first gradually heated to 600°C over a period of five days, keeping the cold end of the ampoule under 250°C. Subsequently, the ampoule was placed in two-zone horizontal furnaces for CVT crystal growth. First, the growth zone was heated to 700°C and the source zone to 550°C. After 2 days, the thermal gradient was reversed, and the growth zone was kept at 600°C and the source zone at 700°C for 15 days. Finally, the ampoule was cooled to room temperature and opened in an argon-filled glovebox.

**Exfoliation of NbOI$_2$.** The NbOI$_2$ samples were exfoliated from a bulk material using the Scotch tape technique to several hundred nm and mechanically transferred by a home-built transfer setup on thermally oxidized silicon substrates (Siltronix, SiO$_2$ thickness 1 mm, roughness < 0.2 nm RMS).

**Growth of SiO$_2$ thin films.** The PVD deposition of SiO$_2$ encapsulation layers was executed in a Syrus Pro 1100© (Bühler Leybold Optics) vacuum evaporation machine at IOF. Pure SiO$_2$ (99,99%) was evaporated in high vacuum conditions, using an electron beam gun. To reach the film thickness of approximately 20nm, a constant deposition rate of 0.4nm/s was realized by quartz crystal monitoring. Ion assistance, as well as plasma pretreatment, which is the standard for the deposition of dielectric optical coatings, was avoided to preserve the sensitive flake[14]. Substrates were cleaned before coating by blowing with ionized air.

**XRD characterization.** The crystal structure and phase identification of NbOI$_2$ were determined using the XRD technique at room temperature by Bruker D8 Discover (Bruker, Germany) with Cu Kα radiation ($\lambda = 1.5418$ Å, I = 40 mA, U = 40 kV). The samples were placed on a silicon plate holder and scanned across the angular spectrum from 5 to 90° with a step size of 0.02° and a step time of 0.5 s. The morphology and elemental composition, as well as the distribution, were determined using scanning electron microscopy (SEM, Tescan Maia3 microscope). The sample was placed on carbon tape and measured using a 10 kV acceleration beam voltage. The energy-dispersive X-ray spectroscopy was determined by an EDS analyzer (X-MaxN) with a 150 mm$^2$ SDD detector (Oxford Instruments). Data were evaluated in AZtecEnergy software.

**XPS characterization.** The surface composition of synthesized NbOI$_2$ was characterized with X-ray photoelectron spectroscopy with a SPECS spectrometer equipped with a monochromatic Al Kα X-ray source (1486.7 eV) and a hemispherical electron analyzer Phoibos 150. The survey spectrum was recorded with E$_p$ set to 100 eV, and the high-resolution spectra of the core lines with E$_p$ set to 40 eV. The base chamber pressure during the acquisitions was at 10$^{-9}$ mbar or lower.

**Raman spectroscopy.** We investigated the crystal structure and morphologies of NbOI$_2$ before and after oxide encapsulation using Raman spectroscopy (Renishaw's inVia confocal Raman microscopes operated in backscattering mode with a 473 nm wavelength, coupled with a 100x objective and a thermoelectrically cooled CCD detector). The spectral resolution of the system is narrower than 0.5 cm$^{-1}$. For all spectra, the Si peak at 520 cm$^{-1}$ was used for peak shift calibration of the instrument.

**Second-harmonic generation.** The second-harmonic generation measurement was performed using a femtosecond laser (Light conversion Carbide, CB3-80W) emitting pulses with a duration of 210 fs and a repetition rate of 2 MHz with a tunable wavelength range of 315–2600 nm through optical parametric amplification (Light conversion Orpheus). The excitation wavelength of 1550 nm with a fixed vertical input polarization and an average power of 0.5 mW was chosen for excitation. The sample was placed on an XYZ piezo controller stage (Physics Instruments). A pair of 50x objectives with an NA of 0.42 was used to focus the fundamental beam on the sample as well as to collect the transmitted second harmonic generation signal. To allow the SH signal into our camera (Zyla 4.2 sCMOS) and block the fundamental beam, we have used several filters (short-pass filters at 800 nm and 950 nm, and long-pass filters at 600 nm and 700 nm).

**Acknowledgements**

The Authors acknowledge BMBF Project SINNER Grant No. 16KIS1792, the Collaborative Research Centre (CRC/SFB) 1375 NOA, the Fraunhofer Attract Grant SILIQUA No. 40-04866, ERC-CZ program (project LL2101) from the Ministry of Education, Youth and Sports (MEYS), and the project Advanced Functional Nanorobots (reg. No. CZ.02.1.01/0.0/0.0/15_003/0000444 financed by the EFRR).

**Competing interests**

The authors declare no competing interests.

**Data Availability**

Data underlying the results presented in this paper are not publicly available at this time but may be obtained from the authors upon reasonable request.

**Supplementary information.** See Supplement S1-S3 for supporting content.

**References**

1. Bernardi, M., Palummo, M., & Grossman, J. C. Extraordinary sunlight absorption and one nanometer thick photovoltaics using two-dimensional monolayer materials. *Nano Lett.* **13**(8), 3664–3670 (2013).
2. Horng, J. et al. Perfect absorption by an atomically thin crystal. *Phys. Rev. Appl.* **14**(2), 024009 (2020).
3. Malard, L. M. et al. Observation of intense second harmonic generation from $MoS_2$ atomic crystals. *Phys. Rev. B Condens. Matter Mater. Phys.* **87**(20), 201401 (2013).
4. Autere, A. et al. Nonlinear optics with 2D layered materials. *Adv. Mater.* **30**(24), 1705963 (2018).
5. Säynätjoki, A. et al. Ultra-strong nonlinear optical processes and trigonal warping in $MoS_2$ layers. *Nat. Commun.* **8**(1), 893 (2017).
6. Woodward, R. I. et al. Characterization of the second-and third-order nonlinear optical susceptibilities of monolayer $MoS_2$ using multiphoton microscopy. *2D Mater.* **4**(1), 011006 (2017).
7. Kumar, N. et al. Second harmonic microscopy of monolayer $MoS_2$. *Phys. Rev. B - Condens. Matter Mater. Phys.* **87**(16), 161403 (2013).

8. Dogadov, O. et al. Parametric nonlinear optics with layered materials and related heterostructures. *Laser Photonics Rev.* **16**(9), 2100726 (2022).

9. Liu, X. et al. Two-Dimensional Nanostructured Materials for Gas Sensing. *Adv. Funct. Mater.* **27**(37), 1702168 (2017).

10. Anichini, C. et al. Chemical sensing with 2D materials. *Chem. Soc. Rev.* **47**(13), 4860–4908 (2018).

11. An, N. et al. Electrically tunable four-wave-mixing in graphene heterogeneous fiber for individual gas molecule detection. *Nano Lett.* **20**(9), 6473–6480 (2020).

12. Tan, T. et al. Multispecies and individual gas molecule detection using Stokes solitons in a graphene over-modal microresonator. *Nat. Commun.* **12**(1), 6716 (2021).

13. Ngo, G. Q. et al. Photoluminescence-based gas sensing with $MoS_2$ monolayers. *Optics Express* **33**(13), 27791–27799 (2025).

14. Ngo, G. Q. et al. Scalable oxide encapsulation of CVD-grown $WS_2$ and $WSe_2$ for photonic applications. *Opt. Mater. Express* **15**(6), 1330–1341 (2025).

15. Radisavljevic, B. et al. Single-layer $MoS_2$ transistors. *Nat. Nanotechnol.* **6**(3), 147–150 (2011).

16. Ovchinnikov, D. et al. Electrical transport properties of single-layer $WS_2$. *ACS Nano* **8**(8), 8174–8181 (2014).

17. Zhou, H. et al. Large area growth and electrical properties of p-type $WSe_2$ atomic layers. *Nano Lett.* **15**(1), 709–713 (2015).

18. Najafidehaghani, E. et al. 1D *p–n* junction electronic and optoelectronic devices from transition metal dichalcogenide lateral heterostructures grown by one-pot chemical vapor deposition synthesis. *Adv. Funct. Mater.* **31**(27), 2101086 (2021).

19. Liu, L. et al. Enhanced optical Kerr nonlinearity of $MoS_2$ on silicon waveguides. *Photonics Res.* **3**(5), 206–209 (2015).

20. Chen, H. et al. Enhanced second-harmonic generation from two-dimensional $MoSe_2$ on a silicon waveguide. *Light Sci. Appl.* **6**(10), e17060 (2017).

21. Zhang, Y. et al. Enhanced four-wave mixing with $MoS_2$ on a silicon waveguide. *J. Opt.* **22**(2), 025503 (2020).

22. Ngo, G. Q. et al. Scalable functionalization of optical fibers using atomically thin semiconductors. *Adv. Mater.* **32**(47), 2003826 (2020).

23. Zuo, Y. et al. Optical fibres with embedded two-dimensional materials for ultrahigh nonlinearity. *Nat. Nanotechnol.* **15**(12), 987–991 (2020).

24. Ngo, G. Q. et al. In-fibre second-harmonic generation with embedded two-dimensional materials. *Nat. Photonics* **16**(11), 769–776 (2022).

25. Li, Y., et al. Nonlinear co-generation of graphene plasmons for optoelectronic logic operations. *Nat. Commun.* **13**(1), 3138 (2022).

26. Gholipour, B. et al. Roadmap on chalcogenide photonics. *J. Phys. Photonics* **5**(1), 012501 (2023).

27. Schneider, L. M. et al. The influence of the environment on monolayer tungsten diselenide photoluminescence. *Nano-Structures and Nano-Objects* **15**, 84–97 (2018).

28. Tyagi, D. et al. Recent advances in two-dimensional-material-based sensing technology toward health and environmental monitoring applications. *Nanoscale* **12**(6), 3535–3559 (2020).

29. Fang, Y. et al. 2D $NbOI_2$: A Chiral Semiconductor with Highly In-Plane Anisotropic Electrical and Optical Properties. *Adv. Mater.* **33**(29), (2021).

30. Abdelwahab, I. et al. Giant second-harmonic generation in ferroelectric $NbOI_2$. *Nat. Photonics* **16**(9), (2022).

31. Guo, Q. et al. Ultrathin quantum light source with van der Waals $NbOCl_2$ crystal. *Nature* **613**(7942), (2023).

32. Tang, T. et al. Third Harmonic Generation in Thin $NbOI_2$ and $TaOI_2$. *Nanomaterials* **14**(5), (2024).

33. Cui, Y. et al. Piezoelectricity in $NbOI_2$ for piezotronics and nanogenerators. *npj 2D Mater Appl* **8**, 62 (2024).

34. Zhang, Z. et al. Giant electro-optic and elasto-optic effects in ferroelectric $NbOI_2$. *Physical Review B*, **110**(10), L100101 (2024).

35. Yan, Q. et al. Ambient Degradation Anisotropy and Mechanism of van der Waals Ferroelectric $NbOI_2$. *ACS Appl. Mater. Interfaces* **16**(7), (2024).

36. Wang, R. et al. Tailoring the Charge Transfer-Driven Oxidation in van der Waals Ferroelectric $NbOI_2$ Through Hetero-Interface Engineering. *Adv. Funct. Mater.* **35**(6), 2414753 (2025).

37. Mortazavi, B. et al. Highly anisotropic mechanical and optical properties of 2D $NbOX_2$ (X= Cl, Br, I) revealed by first-principle. *Nanotechnology* **33**(27), 275701 (2022).

38. Kim, S. et al. Realization of a high mobility dual-gated graphene field-effect transistor with $Al_2O_3$ dielectric. *Appl. Phys. Lett.* **94**(6), 062107 (2009).

39. Vaziri, S., Östling, M. & Lemme, M. C. A hysteresis-free high-k dielectric and contact resistance considerations for graphene field effect transistors. *ECS Trans.* **41**(7), 165−171 (2011).

40. Fallahazad, B. et al. Scaling of $Al_2O_3$ dielectric for graphene field-effect transistors. *Appl. Phys. Lett.* **100**(9), 093112 (2012).

41. Khan, M. F. et al. Electrical and photo-electrical properties of $MoS_2$ nanosheets with and without an $Al_2O_3$ capping layer under various environmental conditions. *Sci. Technol. Adv. Mater.* **17**(1), 166−176 (2016).

42. Liu, H. et al. The integration of high-k dielectric on two-dimensional crystals by atomic layer deposition. *Appl. Phys. Lett.* **100**(15), 152115 (2012).

43. Jia, Y. et al. Niobium oxide dihalides NbOX$_2$: a new family of two-dimensional van der Waals layered materials with intrinsic ferroelectricity and antiferroelectricity. *Nanoscale Horiz.* **4**, 1113–1123 (2019).

44. Rijnsdorp, J.& Jellinek, F. The Crystal Structure of Niobium Oxide Diiodide NbOI$_2$. *J. Less-Common Met.* **61** (1), 79−82 (1978).

45. Yoon, D. et al. Interference effect on Raman spectrum of graphene on SiO$_2$/Si. *Phys. Rev. B* **80**, 125422 (2009).

# Supplementary Information:

# Stabilizing van der Waals NbOI$_2$ by SiO$_2$ encapsulation for Photonic Applications

Gia Quyet Ngo,[1,2,*] Fatemeh Abtahi,[1] Jakub Regner,[3] Hossein Esfandiar,[2] Peter Munzert,[2] Jan Plutnar,[3] Zdeněk Sofer,[3] Falk Eilenberger,[1,2,4] and Sebastian W. Schmitt[1,2]

[1]*Institute of Applied Physics, Abbe Center of Photonics, Friedrich Schiller University Jena, Albert-Einstein-Str. 15, 07745 Jena, Germany*

[2]*Fraunhofer-Institute for Applied Optics and Precision Engineering IOF, Albert-Einstein-Str. 7, 07745 Jena, Germany*

[3]*Department of Inorganic Chemistry, University of Chemistry and Technology, Prague Technicka´ 5, Prague, 616628, Czech Republic*

[4]*Max Planck School of Photonics, Germany*

*quyet.ngo@uni-jena.de

# SUPPLEMENT S1 – Atomic force microscopy of unencapsulated NbOI$_2$

Atomic force microscopy (AFM) measurements were performed by a Dimension Edge AFM (Bruker) using Tap300Al-G tips in tapping mode, a maximum of 25 × 25 μm$^2$ scanning range, a resolution of 512x512 pixels, and a low sampling rate of 0.2 Hz (scanning time per image 42 min). The exfoliated NbOI$_2$ flakes ranged in size from 100 to 200 μm. Because of this, the NbOI$_2$ flakes were only partially mapped by AFM. The scanned range was adjusted for each sample.

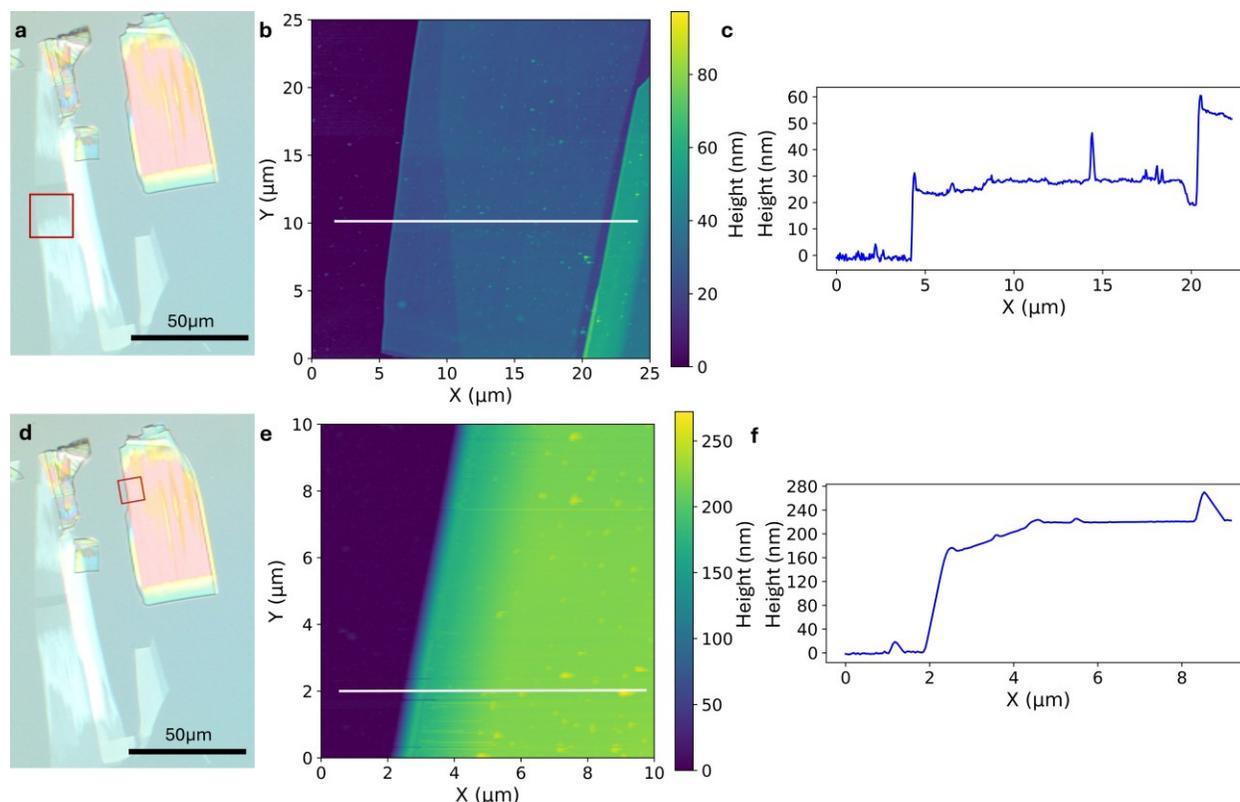

**Figure S1**. Atomic force microscopy of unencapsulated reference NbOI$_2$ flakes. **a**, Bright field microscopy image of the first unencapsulated flake. **b**, AFM topography map of the measured region, marked in the red square box crossing the degraded area in **a**. **c**, AFM topography profile as measured along the white lines in **b**. **d**, Bright field microscopy image of the second unencapsulated flake. **e**, AFM topography map of the measured region, marked in the red square box in **d**. **f**, AFM topography profile as measured along the white lines in **e**.

Figure S1a presents the OM images of the unencapsulated reference NbOI$_2$ flake discussed previously at day 31, to observe the morphology of the degraded area. Fig. S1b shows the topography map of the degraded NbOI$_2$ region marked as the red box. The scan crosses the degraded area to show the change in height and roughness. Fig. S1c displays the height profiles of the degraded flake at the y-coordinate of 10 μm, measured along the white lines in Fig. S1b. The thickness of the reference flake is determined to be in the range of 28-60 nm. Many small islands can be observed in the entire scanned area, which can be linked to the degradation process. Fig. S1d shows the measured area on the second unencapsulated flake. Height profiles and topography for the second unencapsulated flake are presented in Figs. S2e-f, which shows a thickness of 220 nm. Like the first unencapsulated flake, many large islands can be seen in the topography.

**SUPPLEMENT S2 – Atomic force microscopy of encapsulated NbOI$_2$**

Figures S2a,d show the OM images of the encapsulated flake with the measured areas indicated by square boxes. We investigated this flake to determine its protection over time. We can see that two regions of the encapsulated flakes have different contrast, indicating different heights. Fig. S2b shows the topography map of the first region of the encapsulated NbOI$_2$, illustrating a typical surface morphology. The profile line is marked in Fig. S2b for ease of presentation. In Fig. S2b, the height profile of the NbOI$_2$ crystals is clearly visible based on the colour bar. The maximum height can reach 400 nm at the edge. Fig. S2c presents the height profiles of the encapsulated flake, as indicated by the white lines in the AFM topography scans. The measured height at the y-coordinate of 5 µm is 360 nm. Fig. S2e-f shows the thickness of the second area, determined to be around 230 nm. The topography maps of two different scanned areas of the encapsulated NbOI$_2$ reveal a smooth surface of the flake, which contrasts with the unencapsulated flakes.

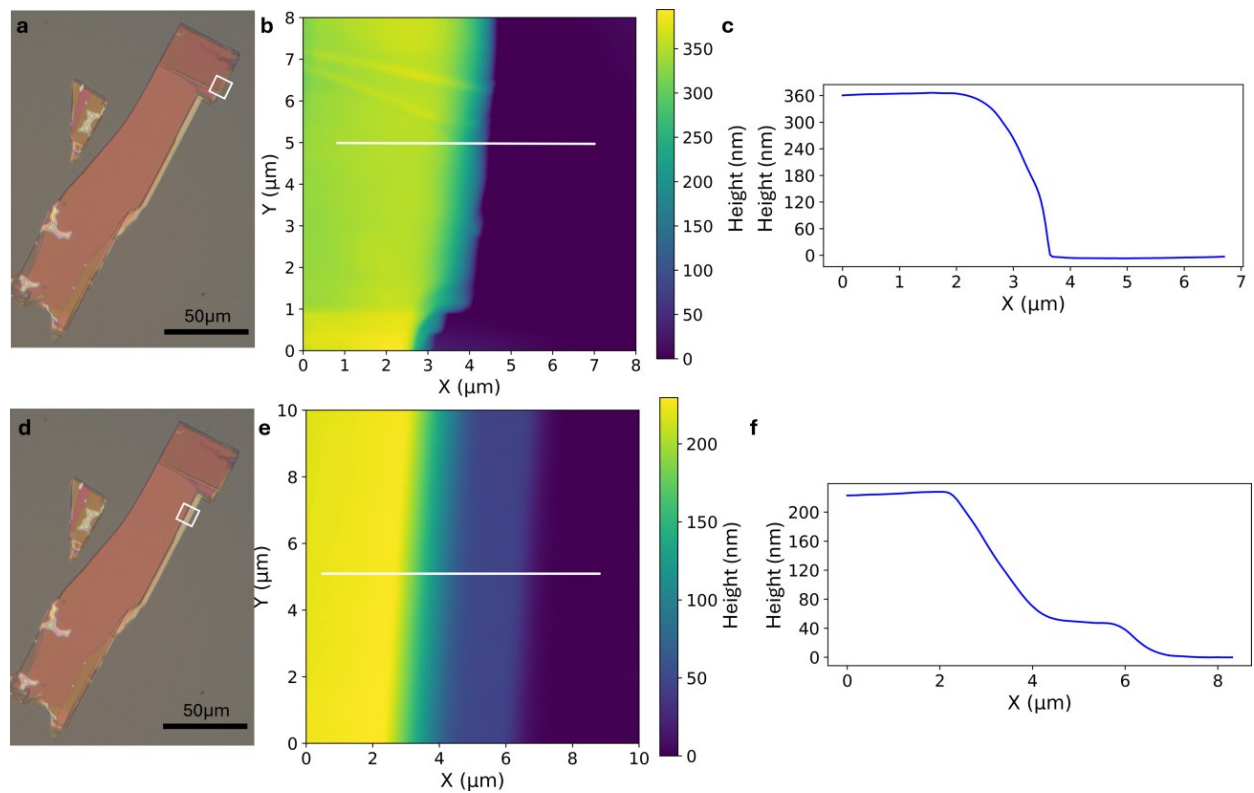

**Figure S2.** AFM topography of encapsulated NbOI$_2$ sample measured at different positions. **a,** Bright field microscopy image of the encapsulated flake with the first measured area. **b,** AFM topography map of the first measured region in **a**. **c,** AFM topography profile as measured along the white lines in **b**. **d,** Bright field microscopy image of the second measured area. **e,** AFM topography map of the second measured region in **d**. **f,** AFM topography profile as measured along the white lines in **e**.

# SUPPLEMENT S3 – Encapsulation of NbOI$_2$ with Al$_2$O$_3$ thin film

In order to test the generality of the finding with respect to different oxides, we have transferred a NbOI$_2$ flake onto a SiO$_2$/Si substrate. Similarly, a 75 nm Al$_2$O$_3$ thin film was deposited using physical vapor deposition on a NbOI$_2$ flake for an encapsulation study. Figures S3a-c show the microscopic image of the sample before and after the encapsulation. The crystallographic morphology of the flake was retained after the coating process. The change of colour before the coating and after the coating is a sequence of thin film interference. The morphology and geometry of the sample remained unchanged at day 15 after the encapsulation, indicating the protection of Al$_2$O$_3$ from the environment. Fig. S3d presents the Raman spectrum of the encapsulated sample using the identical setup and settings as in the Raman spectroscopy of the SiO$_2$ encapsulated sample. There are four distinct Raman peaks observed, like the previous samples in the main text. However, the intensity of those peaks here is more homogeneous than in the case of unencapsulated and SiO$_2$-encapsulated NbOI$_2$. The peak at 520 cm$^{-1}$ originates from the Si substrate.

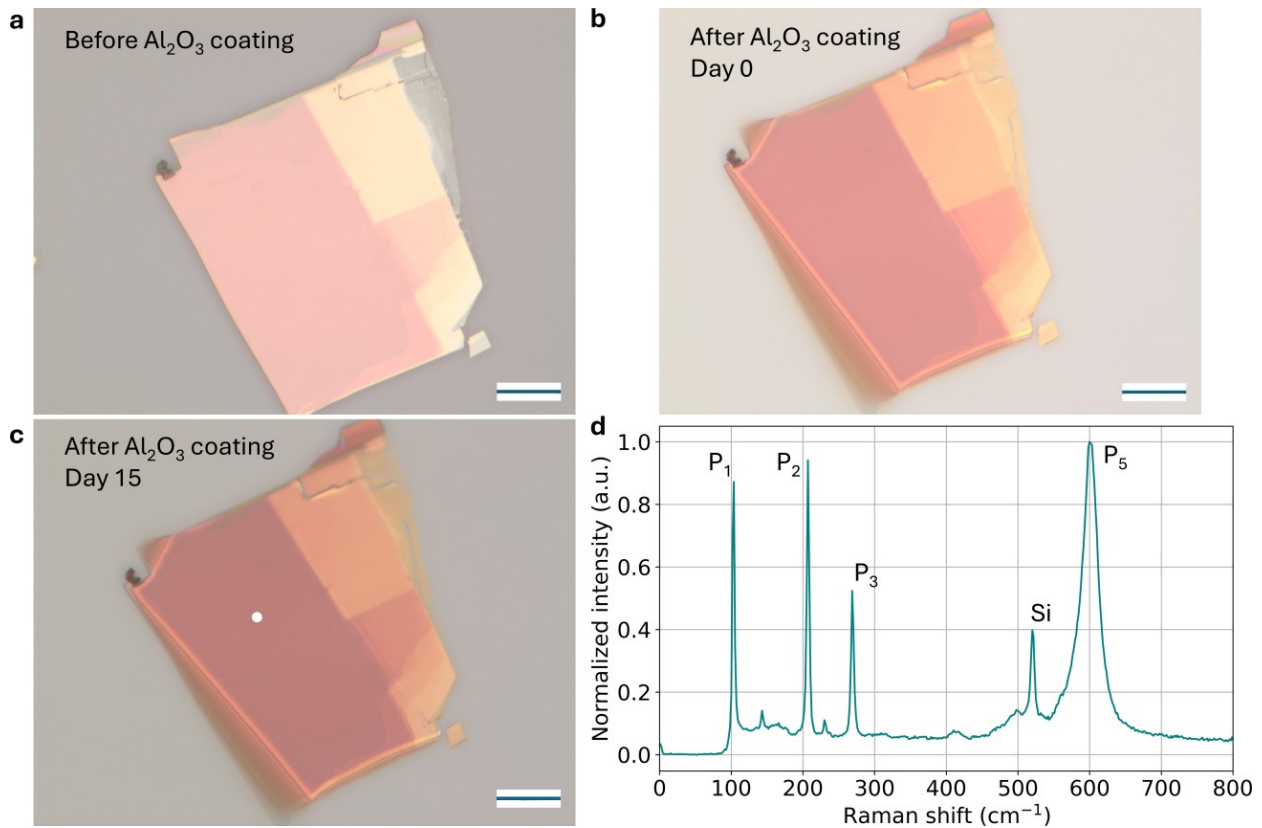

**Figure S3.** Optical microscopic images of NbOI$_2$ before and after encapsulation by a 75 nm Al$_2$O$_3$ film and Raman spectrum. **a**, NbOI$_2$ flake was transferred on SiO$_2$/Si substrate before PVD coating. **b**, Encapsulated NbOI$_2$ right after PVD coating with 75 nm Al$_2$O$_3$. **c**, Encapsulated NbOI$_2$ at day 15. The white dot indicates the measured position of Raman spectroscopy. The scale bar is 20 μm. **d**, Raman spectrum measured from encapsulated NbOI$_2$. P$_1$-P$_5$ are intrinsic Raman peaks from NbOI$_2$ discussed in the main text. The Raman peak at 520 cm$^{-1}$ is from the Silicon substrate.